\documentclass[preprint,3p]{elsarticle}

\usepackage{amssymb}

\usepackage{graphicx}
\usepackage{color}

\newcommand{\solar}{\odot}

\begin{document}

\begin{frontmatter}

\title{Direct dark matter detection: the next decade}
 \author{Laura Baudis}
 \address{Physik Institut, University of Zurich, Winterthurerstrasse 190, 8057 Zurich}
 
\author{}
\address{}

\date{\today}

\begin{abstract} 
Direct dark matter searches are promising techniques to identify the nature of dark matter particles. I describe the future of this field of research, focussing on the question of what can be achieved in the next decade.  I will present the main techniques and R\&D projects that will allow to build so-called ultimate WIMP detectors, capable of probing spin-independent interactions  down to the unimaginably low cross section of 10$^{-48}$cm$^2$, before the irreducible neutrino background takes over. If a  discovery is within the reach of a near-future dark matter experiment, these detectors will be able to constrain WIMP properties such as its mass, scattering cross section and possibly spin. With input from the LHC and from indirect searches, direct detection experiments  will hopefully allow to determine the local density and to constrain the local phase-space structure of our dark matter halo.

\end{abstract}

\end{frontmatter}

\section{Introduction}

The last decades have brought tremendous progress in understanding the structure  of our universe. We have unequivocal evidence that the majority of the material that forms galaxies, clusters of galaxies and the largest observed structures is non-luminous, or dark.  This conclusion rests upon accurate measurements of galactic rotation curves,  measurements of orbital velocities  of individual galaxies in clusters, cluster mass determinations via gravitational lensing, precise measurements of the cosmic microwave background acoustic fluctuations and of the abundance of light elements, and upon the mapping of large scale structures. The first quantitative case for a dark matter dominance of the Coma galaxy cluster was made as early as 1933 by the Swiss astronomer Fritz Zwicky \cite{Zwicky:1933gu}.  Since then, our understanding  of the total amount of dark matter and its overall distribution deepened, but we still lack the answer to the most basic question: what is the dark matter made of? One intriguing answer is that it is made of a new particle, yet to be discovered. Instantly, more questions arise: what are the properties of the particle, such as its mass, interaction cross section, spin and other quantum numbers? Is it one particle species, or many? Is it absolutely stable, or very long-lived? Here  the focus will fall on a particular class of dark matter particle candidates, so-called weakly interacting massive particles (WIMPs)\footnote{QCD axions with masses in the range 1\,$\mu$eV - 10\,meV are well-motivated dark matter candidates \cite{Raffelt:2002zz},  recent results from the axion dark matter experiment (ADMX) can be found in \cite{Asztalos:2011ei}.}. As their name suggests, WIMPs have masses around the electroweak scale, are weakly interacting with baryonic matter, charge and color neutral, and either stable, or with lifetimes comparable or larger than the age of our Universe. While their existence is predicted in most beyond-standard-model particle physics theories, such as supersymmetry, models with universal or warped extra dimensions, Little Higgs theories etc, there is no conclusive, direct evidence for their existence yet. I will treat experimental techniques to directly detect WIMPs via tiny energy deposits when they scatter off atomic nuclei in ultra-sensitive, low-background detectors. These techniques are highly complementary to other avenues towards WIMP detection, such as their production at the LHC, and to indirect detection experiments, aiming at observing annihilation products of dark matter particles such as neutrinos, antiprotons, positrons and gamma rays from galactic regions of increased density. 

\section{Direct detection of WIMPs}

The idea that WIMPs can be detected by elastic scattering off nuclei in a terrestrial detector goes back to Goodman and Witten \cite{Goodman:1984dc}, following the suggestion of Drukier and Stodolsky \cite{Drukier:1983gj} to detect solar and reactor neutrinos by exploiting their elastic neutral-current scattering of nuclei in a detector made of superconducting grains embedded in a non-superconducting material.   The study was extended by Drukier, Freese and Spergel \cite{Drukier:1986tm} to include a variety of cold dark matter candidates, as well as details of the detector and the halo model.  They also showed that the Earth's motion around the Sun produces an annual modulation in the expected signal. On the theoretical side, much progress has been made in refining all aspects entering the prediction of scattering event rates: from detailed cross section calculations in specific particle and nuclear physics models, to refined dark matter halo models that take into account uncertainties in the local WIMP density, in their mean velocity and velocity distribution, as well as in the galactic escape velocity. Progress has been tremendous on the experimental side: in developing new technologies that yield an increasing amount of information about every single particle interaction, in applying these technologies to detectors with masses soon to reach the ton-scale, and  in fighting the background noise such that levels below 1 event per kg and year have now been reached. In this section, I will briefly review predictions for signal event rates and signatures, considering specific input and constraints from particle physics and from astrophysical and cosmological measurements.

\subsection{Prediction of event rates}
\label{sec:event_rates}

The differential rate for WIMP elastic scattering off nuclei can be expressed as:

\begin{equation}
\frac {dR}{dE_R}=N_{N}
\frac{\rho_{0}}{m_{W}}
                    \int_{v_{\rm min}}^{v_{\rm max}} \,d \bold{v}\,f(\bold v)\,v
                     \,\frac{d\sigma}{d E_R}\,,  
\label{eq1}
\end{equation}
where $N_N$ is the number of the target nuclei, $m_W$ is the WIMP mass, $\rho_0$ the local WIMP density in the galactic halo, $\bold v$ and $f(\bold v)$ are the WIMP
velocity and velocity distribution function  in the Earth frame  and ${d\sigma}/{d E_R}$ is the WIMP-nucleus differential cross section. The energy that is transferred to the recoiling nucleus is:

\begin{equation} 
E_R=\frac{p^2}{2 m_N} = \frac{{{m_{\rm r}^2}}v^2}{m_N} (1-\cos \theta),
\end{equation}
where $p$ is the momentum transfer, $\theta$ is the  scattering angle in the WIMP-nucleus center-of-mass frame, $m_N$ is the nuclear mass and $m_{\rm r}$ is the WIMP-nucleus
reduced mass:

\begin{equation}
m_r = \frac{m_N \cdot m_W}{m_N + m_W}.
\end{equation}
The minimum velocity is:
 
 \begin{equation} 
v_{\rm min} = \sqrt{\frac{m_N E_{th}}{2m_{\rm r}^2}}\,, 
\end{equation}
where $E_{th}$ is the energy threshold of the detector, and  $v_{\rm max}$ is the escape WIMP velocity in the Earth reference frame.  
The simplest galactic model assumes a Maxwell-Boltzmann distribution for the WIMP velocity in the galactic rest frame with a velocity dispersion  of $\sigma_v \approx$ 270\,km\,s$^{-1}$ and an escape velocity of v$_{esc} \approx$ 544\,km\,s$^{-1}$.  I will discuss these parameters in more detail in section \ref{sec:astrophysics}.

To provide a simple numerical example, I assume that both the nuclear and WIMP masses are 100\,GeV/c$^2$, and that the mean WIMP velocity relative to the target is $\langle v \rangle $ = 220\,km s$^{-1}$ = 0.75$\times$10$^{-3}$c. The mean energy impinged on the nucleus is:

\begin{equation}
\langle E_R \rangle = \frac{1}{2} m_W \langle v \rangle^2 \sim 30 \,\,{\rm keV}.
\end{equation}
Assuming a local dark matter density of $\rho_0$ = 0.3\,GeV\,cm$^{-3}$, the number density of WIMPs is $n_0$ = $\rho_0$/m$_W$, and their flux on Earth:

\begin{equation}
\phi_0 = n_0 \times \langle v \rangle = \frac{\rho_0}{m_W} \times  \langle v \rangle = 6.6 \times 10^4 \,\,{\rm cm^{-2} s^{-1}}.
\end{equation}
An electroweak-scale interaction will have an elastic scattering cross section from the nucleus of $\sigma_{WN} \sim$10$^{-38}$cm$^2$, leading to a rate for elastic scattering:

\begin{equation}
R \sim N_N \times \phi_0  \times \sigma_{WN} = \frac{N_A}{A} \times  \frac{\rho_0}{m_W}  \times \langle v \rangle \times  \sigma_{WN} \sim {\rm 0.13 \,events  \,\,kg}^{-1} {\rm year}^{-1},
\end{equation}
where $N_N$ is the number of target nuclei, $N_A$ is the Avogadro number and $A$ is the atomic mass of the target nucleus. It can also be expressed as:
 
\begin{equation}
R \sim 0.13  \,\, \frac{{\rm events}}{{\rm kg\,\,year}}  \left[ \frac{A}{100} \times \frac{\sigma_{WN}}{10^{-38} \,{\rm cm^2} } \times \frac{\langle v \rangle}{220\, {\rm km\,s^{-1}}} \times \frac{\rho_0}{0.3\, {\rm GeV cm^{-3}}} \right].
\end{equation}

\subsection{Input from particle and nuclear physics}
\label{sec:particle_physics}

While the WIMP mass and  interaction strength can in principle be theoretically predicted, these quantities are only loosely constrained, with  predictions in various BSM theories  spanning many orders of magnitude, in particular regarding the expected interaction cross section. 
An upper bound of 340\,TeV (240\,TeV) on the mass of a stable Majorana (Dirac) fermion which was once in thermal equilibrium  was derived by Griest and Kamionkowski \cite{Griest:1989wd}, based  on partial-wave unitarity of the S-matrix. The unitarity condition bounds the annihilation cross section in the early Universe, which provides a limit on the relic abundance and the mass of the dark matter particle. 
The WIMP-nucleus speed is of the order of 220\,km\,s$^{-1}$, and  the average momentum transfer is:

\begin{equation}
 \langle p \rangle \simeq m_r \langle v \rangle 
\end{equation}
which is in the range between $\sim$6\,MeV/c -- 70\,MeV/c for values of $m_W$ in the range 10\,GeV/c$^2$ -- 1\,TeV/c$^2$.  Hence the elastic scattering occurs in the extreme non-relativistic limit and the low-energy scattering will be isotropic in the center of mass frame. The de Broglie wavelength corresponding to a momentum transfer of $p$=10\,MeV/c  is:
\begin{equation}
\lambda = \frac{h}{p} \simeq 20\, {\rm fm} > r_0 A^{1/3} = 1.25\,{\rm fm} \,\,A^{1/3}
\end{equation}
which is larger than the diameter of most nuclei, apart from the heaviest ones.  The scattering amplitudes on individual nucleons will then add coherently, and only for heavy nuclei and/or WIMPs in the tail of the velocity distribution coherence losses, typically expressed with a nuclear form factor which is smaller than one, will start to play a role.
In the case of a spin-$1/2$ or spin-$1$ WIMP field, the differential WIMP-nucleus cross section can be expressed as the sum of the spin-independent (SI) and spin-dependent (SD) terms:

\begin{equation}
\frac{d\sigma_{WN}}{dE_R} = \frac{m_N}{2 m_r^2 v^2} \left[ \sigma_{SI} F^2_{SI} (E_R) + \sigma_{SD} F^2_{SD} (E_R) \right],
\end{equation}
where $\sigma_{SI}$ and $\sigma_{SI}$ are the cross sections in the zero momentum transfer limits,  $F_{SI}$ and $F_{SD}$ are the nuclear form factors, that depend on the recoil energy and

\begin{equation}
\sigma_{SI} = \frac{4 m_r^2}{\pi} \left[ Z f_p + (A- Z) f_n \right]^2, 
\end{equation}

\begin{equation}
\sigma_{SD} = \frac{32 m_r^2}{\pi} G_F^2 \frac{J+1}{J} \left[  a_p \langle S_p \rangle + a_n  \langle S_n \rangle \right]^2,
\end{equation}
with $f_p, f_n$ and $a_p, a_n$ being the effective WIMP-couplings to neutrons and protons in the spin-independent and spin-dependent case, respectively.  These can be calculated using an effective Lagrangian of the given theoretical model. They depend on the contributions of the light quarks to the mass of the nucleons and on the quark spin distribution within the nucleons, respectively, and on the composition of the dark matter particle\footnote{I note that $a_p$ and $a_n$ are customary defined as $\propto ({\sqrt(2) G_F})^{-1}$, hence the appearance of $G^2_F$ in the spin-dependent, but not in the spin-independent equation.}. $\langle S_p, S_n\rangle = \langle N |S_{p,n} | N \rangle$ are the expectation values of  total proton and neutron spin operators in the limit of zero momentum transfer, and must be determined using detailed nuclear model calculations.  The nuclear form factor for the coherent interaction is taken as the Fourier transform of the nucleon density and is parameterized as a function of momentum transfer $p$ \cite{Lewin:1995rx}:

\begin{equation}
F^2_{SI}(p) = \left( \frac{3 j_1 (p R_1)}{p R_1}  \right)^2 \exp(-p^2 s^2),
\end{equation}
where $j_1$ is a spherical Bessel function, $s\simeq 1$\,fm is a measure of the nuclear skin thickness and $R_1 = \sqrt{R^2 - 5 s^2}$ with $R \simeq 1.25 A^{1/3}$\,fm. At zero momentum transfer, the nuclear form factor is normalized to unity, $F(0) = 1$.  In the spin-dependent case, the form factor is defined as: 

\begin{eqnarray}
F^2_{SD}(p)=\frac{S(p)}{S(0)},
\end{eqnarray}
where the spin-structure functions are commonly written using the decomposition into isoscalar $a_0 = a_p+a_n$ and isovector $a_1 = a_p-a_n$ couplings:

\begin{eqnarray}
S(p) = a_0^2S_{00}(p) + a_0a_1S_{01}(p) + a_1^2S_{11}(p),
\end{eqnarray}
with the three independent form factors, namely the pure isoscalar term $S_{00}$, the pure isovector term $S_{11}$ and the interference term $S_{01}$. Calculations of the expectation values  $\langle S_p, S_n\rangle$ and of the structure functions $S(p)$ are based on the shell-model with various nucleon-nucleon potentials and truncation schemes of the valence space used in the computation \cite{Engel:1989ix,Ressell:1997kx,Holmlund:2004rv,Toivanen:2009zza,Menendez:2012tm}. Ref. \cite{Menendez:2012tm} uses for the first time chiral effective field theory (EFT) currents \cite{Epelbaum:2008ga,Park:2002yp} to determine the spin-dependent couplings of WIMPs to nucleons.   We can immediately see that the spin-independent interaction cross section depends on the total number of nucleons, while the spin-dependent cross section is in general smaller, and only relevant for odd-even nuclei which have a non-zero spin in their ground state.

\subsection{Input from astrophysics}
\label{sec:astrophysics}

Uncertainties in the WIMP velocity distribution $f(\bold{v})$ and in the local dark matter density $\rho_0 \equiv \rho(R_0 = 8\,kpc)$ will translate into uncertainties in the predicted event rates and ultimately in the inferred scattering cross section and WIMP mass.\footnote{A discussion of recent determinations of relevant astrophysical parameters and their systematic errors, as well as a study of the effects of their uncertainties in direct detection experiments can be found in \cite{Green:2011bv}.}  In the so-called standard halo model (SHM), which describes an isotropic, isothermal sphere of collisionless particles with density profile $\rho(r)\propto r^{-2}$, the velocity distribution is Maxwellian:

\begin{equation}
f(\bold{v}) = \frac{1}{\sqrt{2 \pi}\sigma_v} \exp{\left( - \frac{\bold{v}^2}{2 \sigma_v^2} \right)},
\end{equation}
where the velocity dispersion is related to the local circular speed $\sigma_v=\sqrt{3/2} v_c$ with $v_c \equiv v(r = R_0)$. Since in the SHM the density distribution is formally infinite and the velocity distribution extends to infinity, it has to be truncated at the measured local escape velocity $v_{esc} \equiv v_{esc} (R_0)$, such that $ f(\bold{v}) = 0$ for $v \geq v_{esc}$.  Dark matter particles with speeds larger than $v_{esc}(r) = \sqrt{|\phi(r)|} $, where $|\phi(r)|$ is the potential, will not be gravitationally bound to the galaxy. The parameters used in the SHM  are $\rho_0$ = 0.3\,GeV\,cm$^{-3}$= 5$\times$10$^{-25}$\,g\,cm$^{-3}$ = 8$\times$10$^{-3} M_{\solar}$pc$^{-1}$, $v_c$= 220\,km\,s$^{-1}$ and a local escape speed of $v_{esc}$=544\,km\,s$^{-1}$.
The underlying assumption is that the phase-space distribution of the dark matter has reached a steady state and is smooth, which may not be the case for the Milky Way, in particular at the sub-milliparsec scales probed by direct detection experiments\footnote{The Earth's speed with respect to the galactic rest frame is $\sim 0.7 \times10^{13}$m\,yr$^{-1} \sim 0.22$\,mpc\,yr$^{-1}$.}.  High-resolution, dark-matter-only  simulations of Milky Way-like halos find that the dark matter mass distribution at the solar position is indeed smooth,  with substructures  being far away from the Sun. The local velocity distribution of dark matter particles is likewise found to be smooth, and  close to Maxwellian \cite{Vogelsberger:2008qb}. However, because of their finite resolution, numerical simulations typically probe the dark matter distribution on kpc-scales, and whether the local dark matter distribution consists of a number of streams is still an open issue.  Recent simulation of hierarchical structure formation including the effect of baryons revealed that a thick dark matter disk forms in galaxies, along with the dark matter halo \cite{Read:2008fh,Read:2009iv}. The dark disk has a density of $\rho_{d}/\rho_{0}$ = 0.25$-$1.5 and the kinematics are predicted to follow  the Milky Way's stellar thick disk. At the solar neighborhood, this yields a rotation lag of v$_{lag}$=40$-$50\,km/s with respect to the local circular velocity, and a dispersion of  $\sigma \simeq$40$-$60\,km/s.  These velocities are significantly lower than in the SHM and, should such an additional macroscopic dark matter structure be indeed present in our own galaxy, would have implications  for the expected rates in direct \cite{Bruch:2008rx}  and indirect \cite{Bruch:2009rp} dark matter detection experiments.

\subsection{Predicted signatures}
\label{sec:signatures}

To convincingly detect a WIMP-induced signal, a specific signature from a particle populating  our galactic halo is desirable. As detailed in previous sections, the shape of the recoil energy spectrum depends both on the mass of the target nucleus, and on the WIMP mass: for $m_W \ll m_N$, $E_R \propto m_W^2$ and for $m_W \gg m_N$, the recoil energy spectrum is independent of the WIMP mass. This means that the WIMP mass can be determined most accurately when its mass is comparable to the mass of the target nucleus, and that multiple targets with different $m_N$ can help in providing tighter constraints on m$_W$  \cite{Pato:2010zk}.

The Earth's motion through the galaxy induces both a seasonal variation of  the total event rate \cite{Drukier:1986tm,Freese:1987wu} and a forward-backward 
asymmetry in a directional signal \cite{Spergel:1987kx,Copi:1999pw}.  The annual modulation of the WIMP signal arises because of the Earth's motion 
in the galactic rest frame, which is a superposition of the Earth's rotation around the Sun and the Sun's rotation around the galactic center. Since the Earth's orbital speed $v_{orb}$ is much smaller than the Sun's circular speed, the amplitude of the modulation is small (of the order of $v_{\rm orb}$/$v_c\simeq $ 0.07) and the differential rate in the SHM can be written to a first approximation as: 

\begin{equation}
\frac{dR}{dE_R}(E_R,t) \simeq  \frac{dR}{dE_R}(E_R) \left[ 1 + \Delta (E_R) \cos \frac{2\pi  \left(t - t_0\right)}{T} \right]  
\end{equation}
where T = 1\,year and the phase is $t_0=$ 150\,d. $\Delta E$ becomes negative at small recoil energies, meaning that the differential event rates peaks in winter for small recoil energies, and in summer for larger recoils energies \cite{Savage:2006qr}. The energy at which the annual modulation changes phase is also referred to as the crossing energy. Since its value depends both on the WIMP and the target mass, it can in principle be used to determine the mass of the WIMP \cite{Lewis:2003bv}, requiring however very low experimental energy thresholds.

A stronger signature would be given by the ability to detect the axis and direction of the recoil nucleus. Since the WIMP flux in the  lab frame is peaked in the direction of motion of the Sun, namely towards the constellation Cygnus, the recoil spectrum is peaked in the opposite direction. The WIMP interaction rate as a function 
of recoil energy and angle $\gamma$ between the WIMP velocity and recoil direction in the galactic frame is \cite{Spergel:1987kx}:

\begin{equation}
\frac{d^2R}{dE_R d \cos\gamma} = \frac{\rho_0 \sigma_{WN}}{\sqrt{\pi} \sigma_v} \frac{m_N}{2 m_W m_r^2} \exp\left[-\frac{\left[(v_{orb}^E + v_c) \cos\gamma - v_{min}\right]^2}{\sigma_v^2} \right],  
\end{equation}
where $v_{orb}^E$ is the component of the Earth's velocity parallel to the direction of solar motion.
The forward-backward asymmetry yields a large effect of the order of 
$\mathcal{O}$($v_{orb}^E/v_c)\approx$\,1. Thus fewer events, namely a few tens to a few hundred, depending on the halo model, are needed 
to discover a WIMP signal compared to the case where the seasonal modulation effect is exploited \cite{Copi:1999pw,Copi:2000tv}.
The experimental challenge is to build massive detectors capable of detecting the direction of the incoming WIMP.

\section{Backgrounds}
\label{sec:backgrounds}

Minimizing and characterizing the background noise has posed and will remain a continuous challenge for direct dark matter search experiments. 
So far, the main background sources were the environmental radioactivity including airborne radon and its daughters, radio-impurities in the detector construction and shield material, neutrons from $(\alpha,n)$ and fission reactions (so-called radiogenic neutrons, with energies below 10\,MeV),  cosmic rays and their secondaries, and activation of detector materials during exposure at the Earth's surface\footnote{For an excellent review on low-background techniques for rare event searches, I refer to \cite{Heusser:1995wd}.}.  Background sources intrinsic to the detector materials, muon-induced high-energy neutrons and ultimately neutrinos will start to play an increasingly important role in future experiments. 

\subsection{External backgrounds}

While the hadronic component of the cosmic ray flux is rendered negligible by a few tens of meter water equivalent (m\,w.e.)  overburden, muons are more difficult to attenuate and their energy spectrum is shifted to higher energy with increasing depth\footnote{As an example, the mean measured muon energy at a depth of 3600 m\,w.e. is 273\,GeV \cite{Ambrosio:1995cx}.}.  Muons thus penetrate deep underground and produce high-energy neutrons (so-called cosmogenic neutrons, with energies up to tens of GeV) via negative muon capture (dominant only for shallow depths, below 100\,m\,w.e.), photo-nuclear reactions in associated electromagnetic showers, deep-inelastic muon-nucleus scatters, as well as hadronic interactions of nucleons, pions and kaons. These fast neutrons, when attenuated by rock or shields to MeV energies, can produce keV recoils in elastic scatters in the WIMP target. Thus active veto detectors are necessary to reduce this background, by tagging high-energy deposits by the original muon or its associated cascade. Since absolute muon  rates in underground laboratories are low\footnote{The total muon flux measured at the Gran Sasso laboratory is $\simeq3.4\times 10^{-8}$cm$^{-2}$s$^{-1}$ \cite{Bellini:2012te}, the total muon-induced neutron flux is $\simeq2.7\times 10^{-9}$cm$^{-2}$s$^{-1}$; for a compilation of measurements of the differential muon flux as a function of depth, and predictions of induced neutron fluxes, I refer to \cite{Mei:2005gm}.}, detailed modeling of this background component is necessary to predict background rates of existing and  future experiments. Several works have cross-validated existing codes against one another and  have used data where available  \cite{Araujo:2004rv,Mei:2005gm,Lindote:2008nq}. The neutron yield depends on the atomic weight $A$ of the material via a power law $a\cdot A^{b}$, where $b = 0.76 - 0.82$ for 280\,GeV muons, depending on the used simulation package \cite{Araujo:2004rv,Lindote:2008nq}. This means that neutron production in lead shields is about a factor of 20 higher than in hydrocarbons, a fact which is considered in the design of next-generation experimental shields. The intensity of the muon flux underground has been observed to show a temporal variation, with a modulation amplitude of $\sim$1.3\%, a maximum in summer around the end of june and a period of 366 days \cite{Bellini:2012te}. Considerable attention needs to be paid that such a variation does not mimic a seasonal variation in the event rate as predicted by the motion of the Earth-Sun system through the dark matter halo (see section \ref{sec:astrophysics}). 

Traditionally, a combination of low-Z and high-Z materials are employed to diminish the neutron and gamma fluxes coming from the laboratory walls and outer shield layers, and shield structures are kept under a N$_2$-atmosphere at slight overpressure to suppress the background from airborne radon decays and subsequent $^{210}$Pb plate-out. 
Detectors recently constructed or planned for the future increasingly use large water shields, which passively reduce the environmental radioactivity and muon-induced neutrons, and
act as an active muon water Cherenkov veto at the same time. The efficiency of such a shield to both radiogenic and cosmogenic neutrons can be increased by adding an inner layer of boron loaded liquid scintillator \cite{Wright:2010bu}. Typical underground gamma- and radiogenic neutron fluxes of 0.3\,cm$^{-2}$s$^{-1}$ and  9$\times$10$^{-7}$\,cm$^{-2}$s$^{-1}$, respectively \cite{Haffke:2011fp} are reduced by a factor of 10$^6$ after 3\,m and 1\,m of water shield \cite{Selvi:2011zz}. 

\subsection{Internal backgrounds and cosmogenics}

Most dangerous are interaction from neutrons born in  $(\alpha,n)$- and fission reactions from $^{238}$U and $^{232}$Th decays in detector components in the immediate vicinity of dark matter target materials. The neutron energy spectra and yields are  calculated using the exact composition of these materials and the measured amount of  $^{238}$U and $^{232}$Th (or limits thereof) in each component\footnote{Because often secular equilibrium in the primordial decay chains is lost in processed materials, the  $^{238}$U and $^{232}$Th activities are typically determined via mass spectrometry or neutron activation analysis, while the activities of the late part of these chains, $^{226}$Ra,  $^{228}$Ac, $^{228}$Th, are determined via gamma spectrometry using ultra-low background HPGe detectors \cite{Baudis:2011am}.}.  The neutrons are then transported using detailed Monte Carlo simulations to evaluate the expected number of single-scatter nuclear recoils, which might be difficult to distinguish from a potential WIMP signal \cite{Mei:2008ir}. 

Background sources intrinsic to the WIMP target materials such as $^{39}$Ar, $^{85}$Kr, $^{40}$K, radon diffusion and $^{210}$Pb decays at surfaces that are able to mimic WIMP-induced nuclear recoils provide some of the limiting backgrounds of current and possibly future detectors. As an example, very low impurity levels of $\sim$1\,ppt in natural krypton and $\sim$1\,$\mu$Bq/kg radon need to be achieved by ton-scale noble liquid experiments aiming to probe WIMP-nucleon cross sections down to 10$^{-47}$cm$^2$, with typical expected  background rates from external sources below 1 event per ton of target material and year. 

Decays of long-lived isotopes produced by cosmic rays, in particular neutrons\footnote{Neutrons dominate ($\sim$95\%) the nuclide production at the Earth's surface with protons contributing $\sim$5\% of the neutrons.}, in the target and detector materials during their exposure at the Earth's surface are potential sources of background. The calculations of production rates requires data on the cosmic neutron flux at a given altitude and geomagnetic latitude, along with the cross sections for the production of various isotopes. In general, the measured differential cosmic-ray flux at sea level is parameterized, the variation due to geomagnetic rigidity being around 10\%. Semi-empirical formulas based on available experimental data  are used for the production cross sections \cite{Yin:2010zza}.  Available codes such as Activia \cite{Back:2007kk} and Cosmo \cite{Martoff:1992}  can  also calculate the yields of various isotopes after certain exposure and cooling times. However, only for a few materials such as natural and enriched Ge and Cu the predictions can be compared to actual data \cite{Back:2007kk,Cebrian:2010zz}. More data is need for instance in the case of noble liquids such as Ar
and Xe as well as for many other WIMP targets.
 
 \subsection{Neutrino backgrounds}
 
The so-called  ultimate background will come from the irreducible neutrino flux. Solar pp-neutrinos have low energies, but high fluxes, and will contribute to the electronic recoil background via neutrino-electron scattering at the level of $\sim$10-25\,events/(ton$\times$year) in the low-energy, dark matter signal region of a typical detector. Depending on the detector's  discrimination capabilities between electronic and nuclear recoils, solar pp-neutrinos may thus become a relevant background at cross section of 10$^{-48}$cm$^2$ or lower.  Neutrino-induced nuclear recoils from coherent neutrino-nucleus scatters can not be distinguished from a WIMP-induced signal. The $^{8}$B solar neutrinos can produce up to 10$^3$\,events/(ton$\times$year) for heavy targets \cite{Strigari:2009bq}, however these events are below the energy thresholds of current and most likely also future detectors. Nuclear recoils from atmospheric neutrinos and the diffuse supernovae neutrino background will yield event rates in the range 1- 5\,events/(100\,ton$\times$year), depending on the target material, and hence will dominate measured spectra at a WIMP-nucleon cross section below 10$^{-48}$cm$^2$  \cite{Strigari:2009bq,Gutlein:2010tq}. 

\section{Direct detection techniques and current status}

As we have seen, a dark matter particle with a mass in the GeV$-$TeV range  has a mean momentum of a few tens of MeV and an energy below 100\,keV is transferred to a nucleus in a terrestrial detector. Expected event rates range from  one event to less than 10$^{-3}$\,events per kg detector material and year. To observe a WIMP-induced spectrum, a low energy threshold, an ultra-low background noise and a large target mass are essential. In a given detector, the kinetic energy carried by the scattered nucleus is transformed into a measurable signal, such as ionization, scintillation light or lattice vibration quanta (phonons). The simultaneous detection of two observables  
yields a powerful discrimination against background events,  which are mostly interactions with electrons, as opposed to  WIMPs and neutrons, which  scatter  off nuclei.  Highly granular detectors and/or good timing and position resolution will distinguish localized energy depositions from multiple scatters within the active detector volume and in addition allow to  intrinsically  measure the neutron background. The position resolution leads to the identification of events clustered at the detector surfaces or elsewhere,  which are highly improbable to be induced by WIMPs. It is important to underline that the amount of information per individual event is to be maximized such that potential sources of background noise are unlikely to fake an expected-like signal, namely a single-scatter nuclear recoil. 

\subsection{Solid-state cryogenic detectors}

Cryogenic experiments operated at sub-Kelvin temperatures were traditionally leading the field, given their low energy threshold ($<$10\,keV), excellent energy resolution ($<$1\% at 10\,keV)  and the ability to highly differentiate nuclear from electron recoils on an event-by-event basis.  Their development had been driven by the exciting possibility of performing a calorimetric energy measurement down to very low energies with unsurpassed energy resolution.  Because of the T$^3$-dependence of the heat capacity of a dielectric crystal,  at low temperatures a small energy deposition can significantly change the temperature of the absorber. The change in temperature is measured either after the phonons  reach equilibrium, or thermalize, or when they are still out of equilibrium, or athermal, the latter providing additional information about the location of an interaction in the crystal \cite{Booth:1997at}.  The CDMS \cite{Ahmed:2009zw}, CRESST \cite{Angloher:2011uu} and EDELWEISS \cite{Armengaud:2011cy} experiments are successful implementations of these techniques, operating at the Soudan Laboratory, at the Laboratori Nazionali del Gran Sasso (LNGS) and the Laboratoire Souterrain the Modane (LSM), respectively.  CRESST, in a run with 730\,kg$\times$days exposure, claims to have room for a dark matter signal implying a WIMP mass of 25\,GeV and 12\,GeV for a cross section of 1.6$\times$10$^{-42}$cm$^2$ and 3.7$\times$10$^{-41}$cm$^2$, respectively, as obtained from a maximum likelihood analysis that takes into account known backgrounds \cite{Stodolsky:2012wf}. In contrast, CDMS and EDELWEISS have reached sensitivities down to 3.3$\times$10$^{-44}$cm$^2$ for a 90\,GeV/c$^2$ WIMP in a combined analysis with an effective exposure of 614\,kg$\times$days \cite{Ahmed:2011gh} and do not confirm these findings. 

Germanium ionization detectors operated at 77\,K such as CoGeNT\cite{Aalseth:2011wp} and Texono \cite{Li:2011eb} can reach sub-keV energy thresholds and low backgrounds, but lack the ability to distinguish electronic from nuclear recoils. Pulse shape discrimination is employed to discriminate between surface and bulk  events, and to reject events with incomplete charge collection or due to microphonic noise.  Using data from an 18.5\,kg\,day exposure at the Soudan Laboratory, CoGeNT has claimed evidence for a $\sim$7\,GeV WIMP with a cross section for spin-independent couplings around 10$^{-40}$cm$^2$. This potential low-mass WIMP signal was excluded by dedicated searches using CDMS, EDELWEISS and XENON10 data, and seems to be mostly caused by residual surface events \cite{collar_2012}. It has triggered a flurry of activity on the phenomenological side, giving rise to many new dark matter models for low-mass WIMPs. 

\subsection{Noble liquid detectors}

Liquid noble elements such as argon and xenon offer excellent media for building  non-segmented, homogeneous, compact and self-shielding detectors.  Liquid xenon (LXe) and liquid argon (LAr) are good scintillators and ionizers in response to the passage of radiation, and the simultaneous detection of ionization and scintillation signals allows to identify the primary particle interacting in the liquid.  In addition, the 3D position of an interaction can be determined with sub-mm (in the z-coordinate) to mm (in the x-y-coordinate) precisions in a time projection chamber (TPC). These features, together with the relative ease of scale-up to large masses, have contributed to make LXe and LAr powerful targets for WIMP searches \cite{Aprile:1900zz}. The most stringent limits  for spin-independent couplings come from the ZEPLIN-III \cite{Akimov:2011tj,Akimov:2010zz}, XENON10 \cite{Angle:2007uj,Aprile:2010bt} and XENON100 \cite{Aprile:2011hi,Aprile:2011dd}  experiments, reaching a minimum of  2$\times$10$^{-45}$cm$^2$ at a WIMP mass of 55\,GeV/c$^2$ and an effective exposure of 2323.7\,kg$\times$days for the latest XENON100 run \cite{Aprile:2012nq}. The  XMASS experiment \cite{Moriyama:2011zz}, an 835\,kg (100\,kg fiducial) LXe single-phase detector operated at Kamioka, has taken one year of science data and is expected to report first results later this year.

\subsection{Scintillating crystals}

DAMA/LIBRA, a scintillation experiment using 250\,kg of NaI$(Tl)$ crystals, has observed an annual variation in the single-scatter event rate at low energies with an exposure of 1.17\,ton$\times$years and a statistical significance of 8.9\,$\sigma$ \cite{Bernabei:2010mq}.  If interpreted as due to WIMP-induced nuclear recoils, these results are in strong conflict with upper limits from a variety of other experiments, in particular with recents results from KIMS, derived from an exposure of 25\,ton$\times$days with 103.4\,kg CsI$(Tl)$ detectors at the Yangyang underground laboratory \cite{Kim:2012rz}. Projects such as ANAIS \cite{Amare:2011zz} at LSC and DM-Ice  at the South Pole \cite{Cherwinka:2011ij} will also look for this effect in NaI$(Tl)$ crystals.  DM-Ice, which is to operate a 250\,kg experiment at a depth of 2450\,m in the Antarctic icecap, should be able to discern whether the observed modulation is related to cosmic muons, which show a periodic variation with opposite phase as in the Northern Hemisphere.  DAMA/LIBRA will further investigate the origin of the observed variation using a new data set with reduced backgrounds and lower thresholds due to photomultipliers (PMTs) with higher quantum efficiencies \cite{Bernabei:2012zzb}.
 
\subsection{Superheated liquid detectors}

Investigation of the spin-dependent channel requires target nuclei with uneven total angular momentum. A particularly favorable candidate is $^{19}$F, the spin of which is carried mostly by the unpaired proton, yielding a cross section which is almost a factor of ten higher than of other used nuclei, such as $^{23}$Na, $^{73}$Ge, $^{127}$I, $^{129}$Xe and $^{131}$Xe. Fluorine is in the target of WIMP detectors using superheated liquids, such as PICASSO (C$_4$F$_{10}$) \cite{Archambault:2012pm}, COUPP (CF$_3$I) \cite{Behnke:2012ys} and SIMPLE (C$_2$ClF$_5$) \cite{Felizardo:2012ei}.  An energy deposition can destroy the metastable state, leading to the formation of bubbles, which can be detected and recorded both acoustically and optically.  Since a minimal energy deposition is required to induce a phase-transition, these detectors are so-called threshold devices.  The operating temperatures and pressure can be adjusted such that only nuclear recoils (large stopping powers $dE/dx$) 
lead to the formation of bubbles, making them insensitive to electron recoils coming from gamma interactions.  
Alpha-decays from radon and its progenies are a potential problem, and the current generation of experiments is attempting 
to minimize these to negligible levels.  They can be discriminated from nuclear recoils using the acoustic signal, as discovered by PICASSO.  Current detectors have active masses ranging from 0.2\,kg (SIMPLE), to 2.7\,kg (PICASSO), to 4.0\,kg (COUPP), with a 60\,kg  CF$_3$I bubble chamber under installation at SNOLAB. The threshold detectors yield some of the best direct detection limits on spin-dependent WIMP-proton cross sections, namely 5.7$\times$10$^{-39}$\,cm$^2$ at a WIMP mass of 35\,GeV/c$^2$ in the case of SIMPLE.\footnote{The best constraints on  spin-dependent WIMP-neutrons interactions come from CDMS ($^{73}$Ge),  ZEPLIN-III, XENON10 and XENON100 ($^{129}$Xe, $^{131}$Xe) with the lowest probed cross section around 4$\times$10$^{-40}$\,cm$^2$ for a 50\,GeV/c$^2$ mass \cite{Aprile:2011hj}.}

\section{The next decade}

Today, we have no convincing evidence of a direct detection signal induced by galactic WIMPs. Considering XENON100's lack of a signal in 225 live days$\times$34\,kg of liquid xenon target, excluding $\sim$50\,GeV WIMPs with interaction strengths above  $\sim$2$\times$10$^{-45}$cm$^2$, it becomes clear that, at the minimum, ton-scale experiments are required for a discovery above the 5-sigma confidence level\footnote{Unless the WIMP is lighter than $\sim$10\,GeV, where larger cross sections are, in principle, still feasible.}. It is encouraging to see that several large-scale direct detection experiments are in their construction phase and will start taking science data well within this decade. I will briefly discuss these projects, show their relative strengths, and comment on the complementarity of using different materials and techniques. 

\subsection{Solid-state cryogenic detectors}

The European Underground Rare Event Calorimeter Array (EURECA) \cite{Kraus:2011zz}, a collaboration between CRESST, EDELWEISS and new members, proposes to build a 1\,t  cryogenic dark matter detector at the LSM, in a multi-target and phased approach.  The WIMP target materials are cryogenic Ge detectors and scintillating calorimeters, operated in an ultra-pure Cu cryostat surrounded by polyethylene and a 3\,m water shield equipped with PMTs. The first step in this direction is EDELWEISS-III, a major upgrade of both the current cryostat and the Ge detectors.   Forty so-called FID detectors (800\,g each) with rings of charge electrodes on all surfaces for a dramatically improved rejection of surface-events are to be installed at Modane by the end of 2012. With a fiducial mass of $\sim$24\,kg, the goal is to reach a sensitivity of  5$\times$10$^{-45}$cm$^2$ after an exposure of 3000\,kg\,days \cite{Armengaud:2012ef}. 
The conceptual design of EDELWEISS is currently under study;  in its first phase, it will operate a 150\,kg detector array with  a sensitivity goal for the SI cross section of 3$\times$10$^{-46}$cm$^2$ after one year of operation. 

The SuperCDMS experiment \cite{Brink:2012zza} uses new and improved CDMS-style Ge detectors, so-called iZIPs. With an interleaved charge electrode design, the iZIPs yield more than two orders of magnitude higher surface-event rejection than the previous CDMS design, while keeping a larger fiducial volume. Five towers with three iZIPs each, for a total (fiducial) mass of 9\,kg (6\,kg) are taking science data at Soudan. The expected sensitivity is $\sim$5-8$\times$10$^{-45}$cm$^2$ after two years of operation. The next phase foresees the installation of 200\,kg of Ge iZIPs at SNOLAB, with the start of construction in 2014. The physics aim is a sensitivity below 10$^{-46}$cm$^2$ for a 60\,GeV WIMP after 4 years of operation. The phase following SuperCDMS, GEODM,  foresees a ton-scale experiment in a joint effort with EURECA, aiming for a sensitivity of 2$\times$10$^{-47}$cm$^2$.

\subsection{Noble liquid detectors}

Several single- and two-phase noble liquid detectors are under commissioning or construction. LUX is a 350\,kg (100\,kg) total (fiducial) LXe TPC in a water Cherenkov shield at the Sanford Underground Research Facility (SURF) in the Homestake mine. After a successful demonstration that all subsystems are operational at the surface, LUX is installed underground and expects to start a science run at the end of 2012. The sensitivity aim is 7$\times$10$^{-46}$cm$^2$ for a 100\,GeV WIMP after 300 live days. The next phase, LZ, a joint collaboration with ZEPLIN, plans a 7\,t LXe detector in the same SURF infrastructure, with an additional scintillator veto to suppress the neutron background. Construction is expected to start in 2014, and operation in 2016, with the goal of reaching  2$\times$10$^{-48}$cm$^2$ in three years. The XENON1T experiment, a 3\,t (1\,t) total (fiducial) LXe TPC will start construction at LNGS in 2013. To be operated in a 10\,$\times$10\,m water Cherenkov shield and using a design similar to XENON100, XENON1T is to start a first physics run in 2015. It aims to reach a sensitivity of 2$\times$10$^{-47}$cm$^2$ after two years of operation underground. The XMASS collaboration plans a 5\,t (1\,t fiducial) single-phase detector after its current phase, to start operation in 2015 with a sensitivity of 10$^{-46}$cm$^2$. On the liquid argon side, DarkSide \cite{Wright:2011pa}, a 50\,kg active mass, two-phase LAr experiment is under construction at LNGS, while ArDM, a TPC with a total of 850\,kg of LAr \cite{Marchionni:2010fi}, is under commissioning at LSC.  At SNOLAB, two single-phase argon detectors are under construction: MiniCLEAN, with 180\,kg of LAr in the fiducial volume, and DEAP3600, which will operate 3.6\,t (1\,t) total (fiducial) mass of LAr in a spherical geometry surrounded by an 8\,m water Cherenkov shield \cite{Boulay:2012hq}. While the detector construction will start at the end of 2012, data taking is expected to start one year later, with an aimed sensitivity reach of $\sim$10$^{-46}$cm$^2$ for a 100\,GeV WIMP. 

\begin{figure}[!h]
\centering
\includegraphics[scale=0.50]{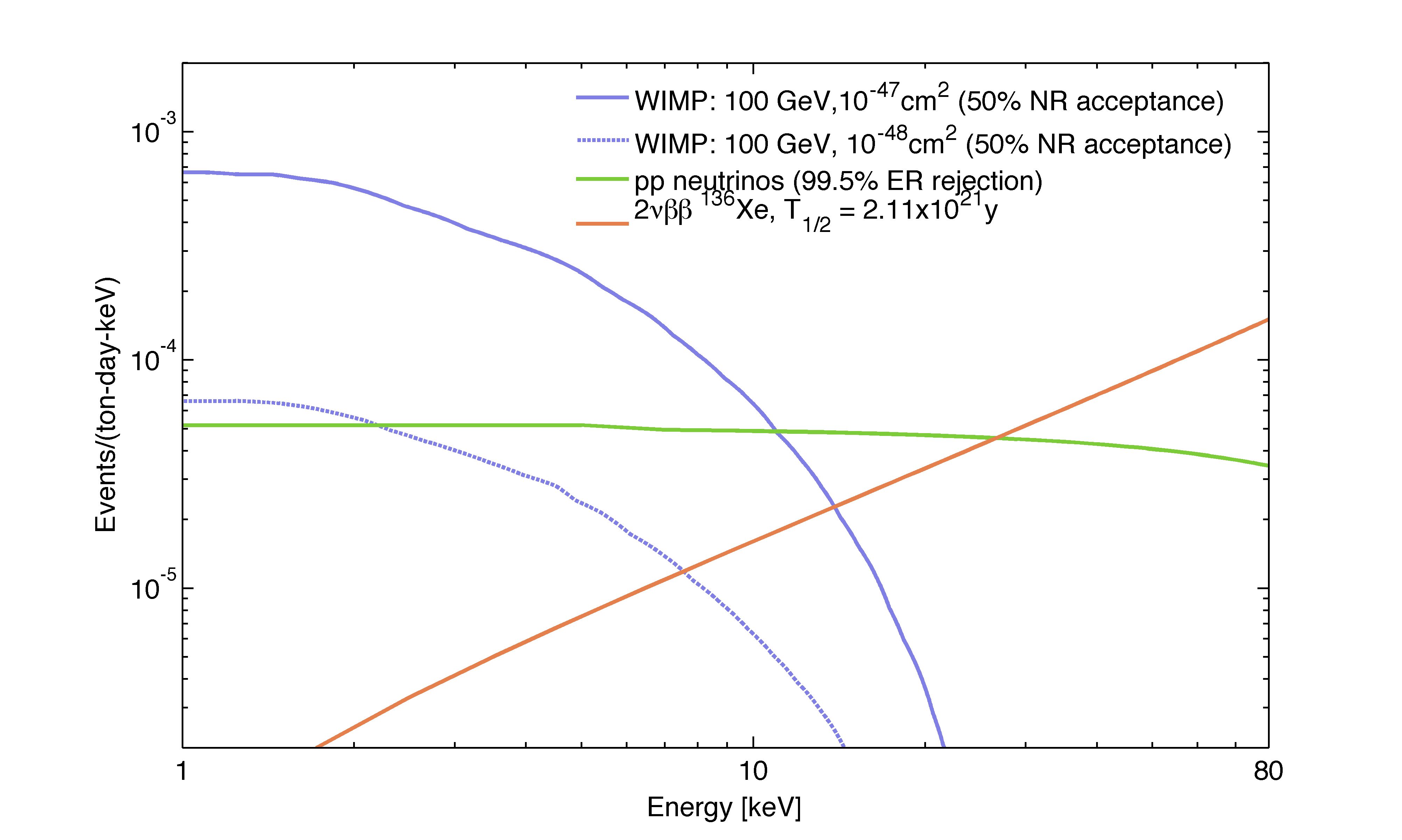}
\caption{\small{Expected nuclear recoil spectrum from WIMP scatters in LXe for spin-independent WIMP-nucleon cross sections of 10$^{-47}$\,cm$^2$ and 10$^{-48}$\,cm$^2$  and a WIMP mass of 100\,GeV/c$^2$, along with the differential energy spectrum  for solar pp-neutrinos, and the electron recoil spectrum from the double beta decay of $^{136}$Xe, assuming the natural abundance of 8.9\% and the recently measured half life of  2.1$\times$10$^{21}$\,yr \cite{Ackerman:2011gz}.  Other assumptions are: 99.5\% discrimination of electronic recoils, 50\% acceptance of nuclear recoils.}}
\label{fig:darwin_wimp_neutrinos}
\end{figure}

Looking further into the future, two large-scale noble liquid programs, using LAr and LXe two-phase detectors are under study. MAX, in the US, with 70\,t (40\,t) and 20\,t (10\,t) total (fiducial) LAr and LXe detectors, respectively, surrounded by liquid scintillators and 15\,m diameter water tanks at DUSEL \cite{Alarcon:2009zz}.  DARWIN (Dark matter WIMP search with noble liquids) \cite{Baudis:2010ch, Baudis:2012bc} is a European design study for 20\,t (10\,t)  LXe and/or LAr TPCs operated in large water Cherenkov shields either at LNGS, or in the planned LSM extension.  Both studies build upon the acquired and near-future experience with large LAr/LXe dark matter detectors and have extended physics programs such as the detection of low-energy solar neutrinos and of the neutrinoless double beta decay in $^{136}$Xe.  Their sensitivity to coherent WIMP-nuclei interactions will be in fact limited by the irreducible solar neutrino flux,  yielding a lower limit on the attainable cross section of  $\sim$10$^{-48}$cm$^2$ due to elastic neutrino-electron scattering of pp-neutrinos, for an assumed discrimination of electronic versus nuclear recoils of 99.5\% and an acceptance of nuclear recoils of 50\%, as shown in Figure \ref{fig:darwin_wimp_neutrinos}.

\subsection{Directional detectors}

Detectors capable of measuring the direction of the recoiling nucleus would unequivocally confirm the Galactic origin of a signal. As we have seen, the recoil spectrum would be peaked in the opposite direction to the constellation Cygnus and ideally a detector would be capable to measure the axis and sense of a WIMP-induced nuclear recoil. A strong signature would require only a few tens of events and terrestrial, seasonal modulations would be unable to fake a potential signal. Because nuclear recoils have a range which is about 10 times smaller than Compton recoils of the same energy, gaseous detectors have an excellent intrinsic background rejection if they can measure the range of events precisely. Several directional detectors are presently in R\&D phase \cite{Ahlen:2009ev}: DRIFT in the Boulby Mine, DMTPC at the Waste Isolation Pivot Plant (WIPP), MIMAC at LSM and NEWAGE in the Kamioka laboratory.  A 1\,m$^3$  detector has a typical mass of a few 100\,g, depending on the target gas and its operating pressure, and can measure the sense of an incoming nuclear recoil above a few tens of keV. These prototypes lay the foundation for the construction of future, much larger directional detectors which are required by the existing constraints on WIMP-nucleon cross sections.

\subsection{Complementarity}

Once substantial evidence for a dark matter signal has been established with so-called discovery experiments, the efforts will shift towards measuring the properties of WIMPs and possibly their local distribution, along with the velocity profile of our dark halo. To reconstruct their physical properties such as mass and cross section with a certain accuracy, based on statistical considerations alone, exposures of several ton-years and multiple targets are required even for a SI cross section as large as 10$^{-45}$cm$^2$, increasing to several tens of ton-years for a cross section of 10$^{-46}$cm$^2$ \cite{Strege:2012kv}.   If we consider a 50\,GeV WIMP as an example, its mass can be reconstructed with a 1-$\sigma$ accuracy of about 5\% when using a combination of data from xenon, germanium and argon experiments and assuming fixed astrophysical parameters. Allowing for 1-$\sigma$ uncertainties in the local density, circular and escape velocity of 0.1\,GeV\,cm$^{-3}$, 30\,km\,s$^{-1}$ and 33\,km\,s$^{-1}$, respectively, increases the 1-$\sigma$ accuracy of the mass determination to about 10\% \cite{Pato:2010zk}. If the astrophysical parameters are left to vary in a broad range, data from dark matter experiments alone, when using multiple targets, can constrain the local circular velocity at least as accurately as it is currently measured  \cite{Pato:2010zk}.

\subsection{Evolution}

\begin{figure}[!h]
\centering
\includegraphics[scale=0.45]{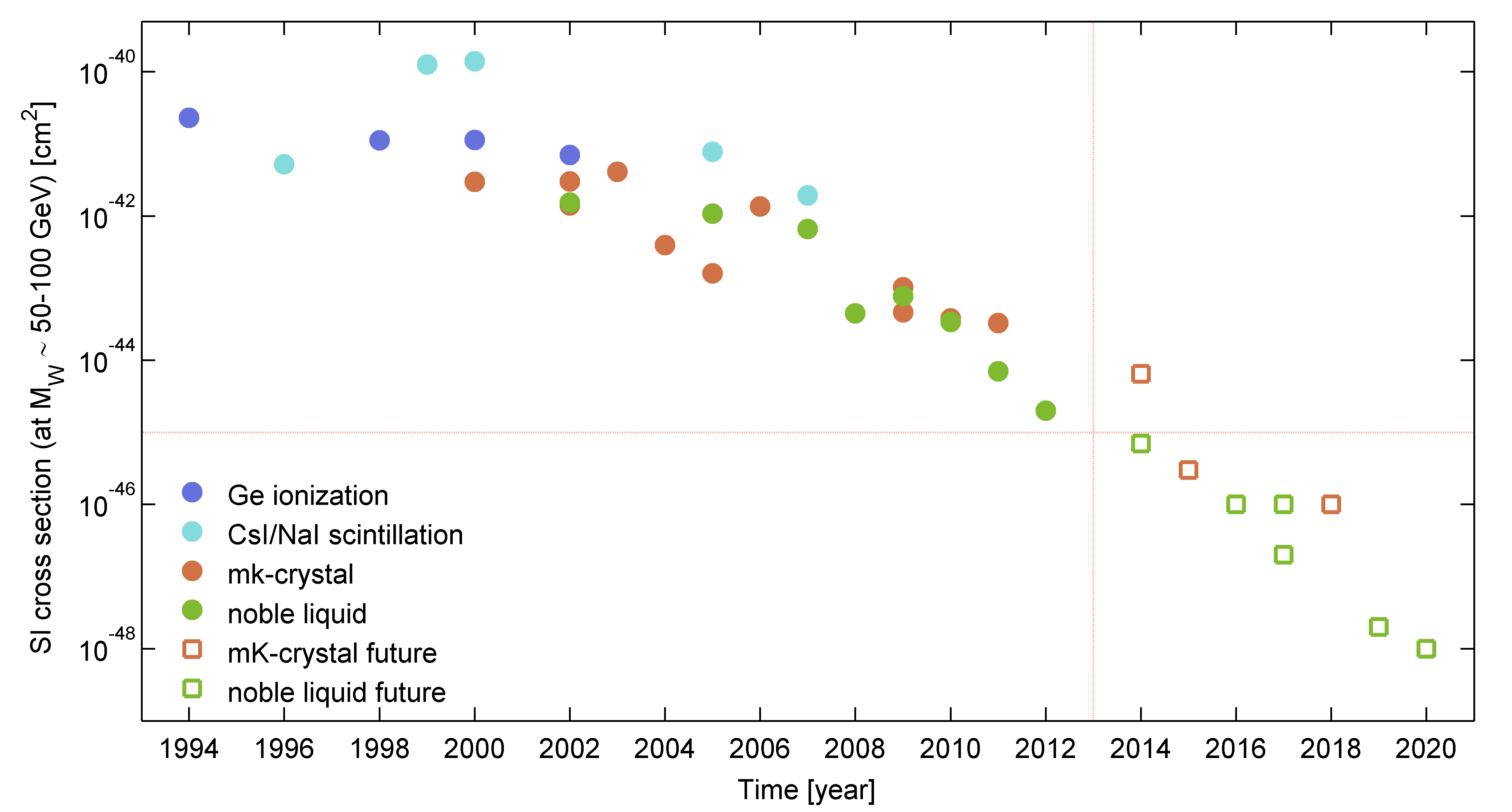}
\caption{\small{Existing upper limits on spin-independent WIMP-nucleon cross sections (for WIMP masses around 50-100\,GeV) from various direct detection techniques (filled circles), along with projections for the future (open squares), as a function time. Part of the data was retrieved from dmtools \cite{dm-tools}. }}
\label{fig:si_limits_time}
\end{figure}

An overview of the evolution of upper limits on the spin-independent cross section as a function of time, together with projections for the future, is shown in Figure\,\ref{fig:si_limits_time}. While the rate of progress was slower during the first decade shown here\footnote{Note however the logarithmic y-scale.}, the rate increased with the advent of cryogenic mK-detectors and of homogeneous noble-liquid detectors capable of fiducialization,  of electronic versus nuclear recoil discrimination and operating deep underground.  Looking at the recent progress, from $\sim$2004-2012, one notices that: i) the sensitivity increased by about one order of magnitude roughly every two years\footnote{Although this trend might be reminiscent of Moore's law, which predicts that the number of transistors on integrated circuits doubles approximately every two years, the rate of sensitivity increase in direct detection experiment is steeper and not related to the increase in computing power. }  ii) the predictions of reachable sensitivities until 2020 seem reasonable if the current trend is extrapolated into the future, assuming however not only larger and lower background detectors, but that known backgrounds can be reduced at the same rate as demonstrated during the last few years.

\section{Epilogue}

A major effort to detect the tiny energy depositions when a galactic WIMP scatters off a nucleus in an ultra-low 
background detector is underway. After decades of technological developments, experiments operating 
deep underground have reached sensitivities to probe and cut deeply into the predicted parameter space of beyond-standard-model particle physics theories. 
The major questions  have shifted from 'how to detect a WIMP' to 'how can we identify its nature' and 'what can we learn about our local, dark 
environment' in case of a clear signal. 

While recent best upper limits on WIMP-nucleon cross sections are derived from  experiments using a few tens of kg target material, larger,  
ton-size detectors are already under operation or construction. These experiments will have a non-negligible chance of 
discovering a dark matter particle and their results will strongly influence the design and implementation of next-generation 
detectors.  The R\&D for multi-ton dark matter experiments, as well as for directional detectors, is well underway.  Considering the lessons we have learned over the past fifty years from the sister field of solar neutrino physics, it seems obvious that large detectors with unimaginable low backgrounds and low energy thresholds must be built to observe dark matter particle interactions with sufficiently high rates to infer their mass and scattering cross section.
In the next decade, a combination of different detector materials and techniques, coupled to using target nuclei with and without spin, will allow - in case of a positive detection  - to 
determine the particle mass, spin, and in some cases, to distinguish among different underlying theoretical models.
Complementary information from dark matter searches at the LHC and from indirect detection experiments such as AMS, Fermi and IceCube, to name a few,  will hopefully allow to determine the local 
density and to constrain the local phase-space structure of our dark matter halo.

 \appendix
 
 \section*{Acknowledgments}
This work is supported by the University of Zurich and by the Swiss National Foundation (SNF) Grants  No 200020-138225, No 20AS21-129329 and  No 20AS21-136660.

\bibliographystyle{elsarticle-num}
\bibliography{baudis_darkuniverse}

\end{document}